# Experimental study of the wind effect on the focusing of transient wave groups

## J.P. Giovanangeli[1], C. Kharif[1] and E. Pelinovsky[1,2]

[1] Institut de Recherche sur les Phénomènes Hors Equilibre, Laboratoire IRPHE/IOA, Marseille, France. Email: giovanangeli@irphe.univ-mrs.fr
[2] Laboratory of Hydrophysics and Nonlinear Acoustics, Institute of Applied Physics, Nizhny Novgorod, Russia. Email: Pelinovsky@hydro.appl.sci-nnov.ru

One of the popular mechanisms of the freak waves phenomenon is the dispersive focusing of transient wave groups. In all published theoretical and experimental papers the surface gravity waves are considered as an ensemble of free waves. This paper reports on a series of experiments conducted in the large wind-wave tank of IRPHE (Marseille – Luminy) to study the wind effect on the generation of freak waves. A suitable theory is presented to explain and discuss the experimental results.

## 1. Introduction

Freak wave phenomenon is now explained by many physical theories: dispersive and geometrical focusing, nonlinear modulational instability (Benjamin – Feir instability), wave-current and wave-bottom interactions; see, for instance (Olagnon and Athanassoulis, 2001; Rogue Waves, 2003; Kharif and Pelinovsky, 2003). In all published theoretical and experimental papers the gravity waves on sea surface are assumed to be an ensemble of free waves. In early stages the wind flow is considered as the source of the spatial and temporal inhomogeneity of the wind wave field or as the factor of wave amplification. The present paper reports on a series of experiments with transient wave groups under wind action up to 10 m/s conducted in the large wind-wave tank of IRPHE (Marseille – Luminy). Theoretical model to explain the experimental results is developed. It includes the sub-surface current induced by wind. The theoretical predictions are in good agreement with experimental results.

## 2. Set-up and experimental conditions

The experiments have been conducted in the large wind-wave tank of IRPHE at Marseille - Luminy (Figure 1). It is constituted of a closed loop wind tunnel located over a water tank 40m long, 1 m deep and 2.6 m wide. The wind tunnel over the water flow is 40 m long, 3.2 m wide and 1.6 m high. The blower allows to produce wind speeds up to 14 m/s and a computer-controlled wave maker submerged under the upstream beach can generate regular or random waves in a frequency range from 0.5 hz to 2 Hz. Particular attention has been taken to simulate pure logarithmic mean wind profile with constant shear layer over the water surface. A trolley installed in the test section allows to locate probes at different fetches all along the facility. The water surface displacements were determined by using three capacitive wave gauges of 0.3 mm outer diameter with DANTEC model 55E capacitance measuring units. Two wave gauges was located at fixed fetches of 1m and 3 m from the uspstream beach. A third wave gauge was installed on the trolley in order to determine the water surface deflections η at various distances from the upstream beach. The typical sensitivity of the wave probes was of order of 0.6 V/cm.

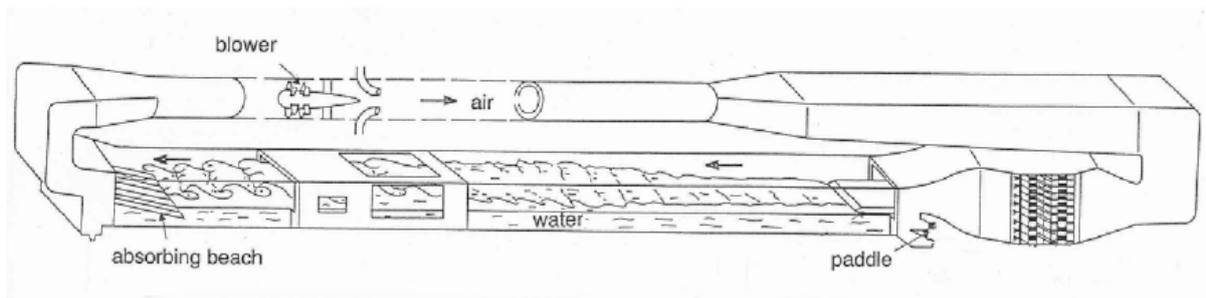

**Figure 1:** A schematic representation of the Large Air-Sea Interactions Facility.

For each value of the mean wind speed, *W* equals to 0, 4, 5, 6, 8 and 10 m/s, the water surface deflections η was measured at 1 m fetch and at different fetches between 5m and 35 m. The wave maker was driven by an analog electronic signal varying linearly with time from 1.3 Hz to 0.8 Hz in 10 seconds with constant amplitude of displacements corresponding to nearly constant amplitude of the initial wave group. The fetch values are taken from the entrance of the wave-tank, where the airflow meets the water surface i.e. at the end of the upstream beach. Typical value of σ equal to the rms of the water deflections η determined at the distance 1 m for different wind speeds is 1.87-1.88 cm.

3. **Experimental Results**

Figure 2 demonstrates the focusing of the free transient wave groups when there is no wind flow above waves. Initial wave packet has the step-wise amplitude modulation and linearly frequency modulation. As predicted by the linear theory of the free deep-water waves, the waves focused at a precise distance, leading to the occurence of a high amplitude freak wave. Downstream of the point of focusing, the amplitude of the group decreases rapidly (defocusing). The influence of the wind flow with speed 6 m/s on the wave focusing process is shown in Figure 3. The characteristics (frequencies and amplitude) of the initial mechanically generated waves are kept the same. As it can be seen, the scales of spatial - temporal evolution of the group are changed with wind. For each value of the wind speed, the amplification ratio *A(X,W)* can be defined as $A = \eta_{max}/\sigma$, where $\eta_{max}$ is the maximum amplitude of the wave packet on fixed distance. Figure 4 gives the amplification ratio as a function of the distance from the upstream probe located at 1 m fetch and for different values of the wind speed. This figure shows that in presence of wind, we observed a freak wave of a larger amplitude, occuring at a longer distance than observed without wind. Moreover, contrarely to the case without wind, the freak waves maintains its coherency and its amplitude as it propagates downstream the previous focusing point. The effects of wind on the location of the focusing point, on the amplitude and the coherency of the freak wave increases as the wind velocity increases.

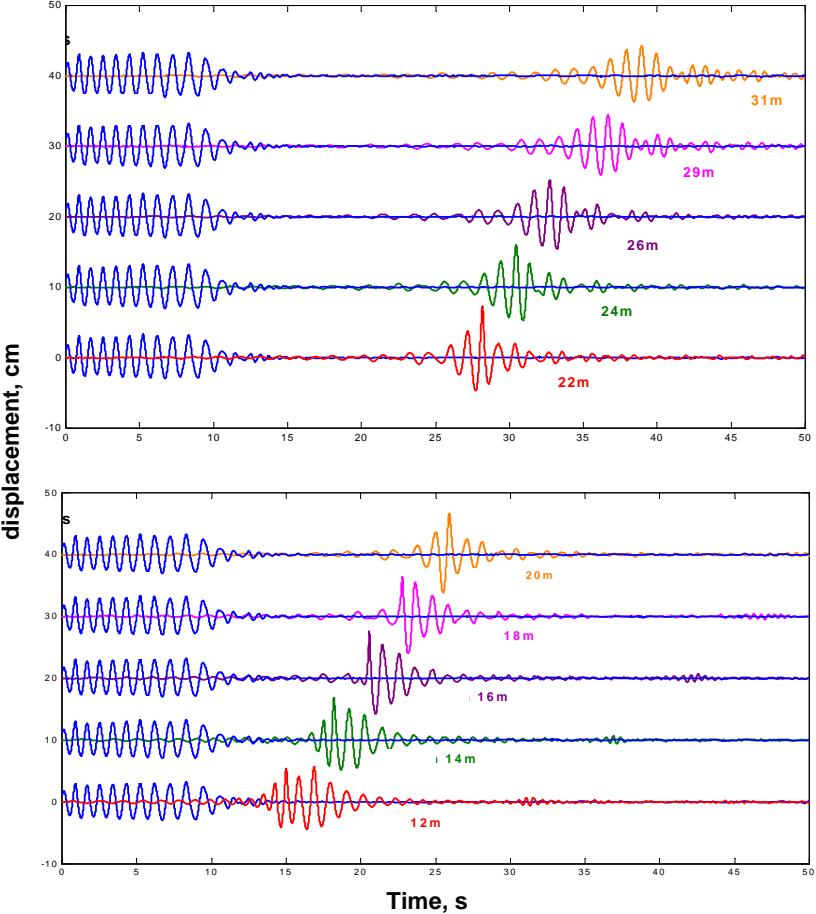

**Figure 2:** Time series of the water surface displacement, η on various distances from the wave paddle (no wind); initial packet is shown on the left

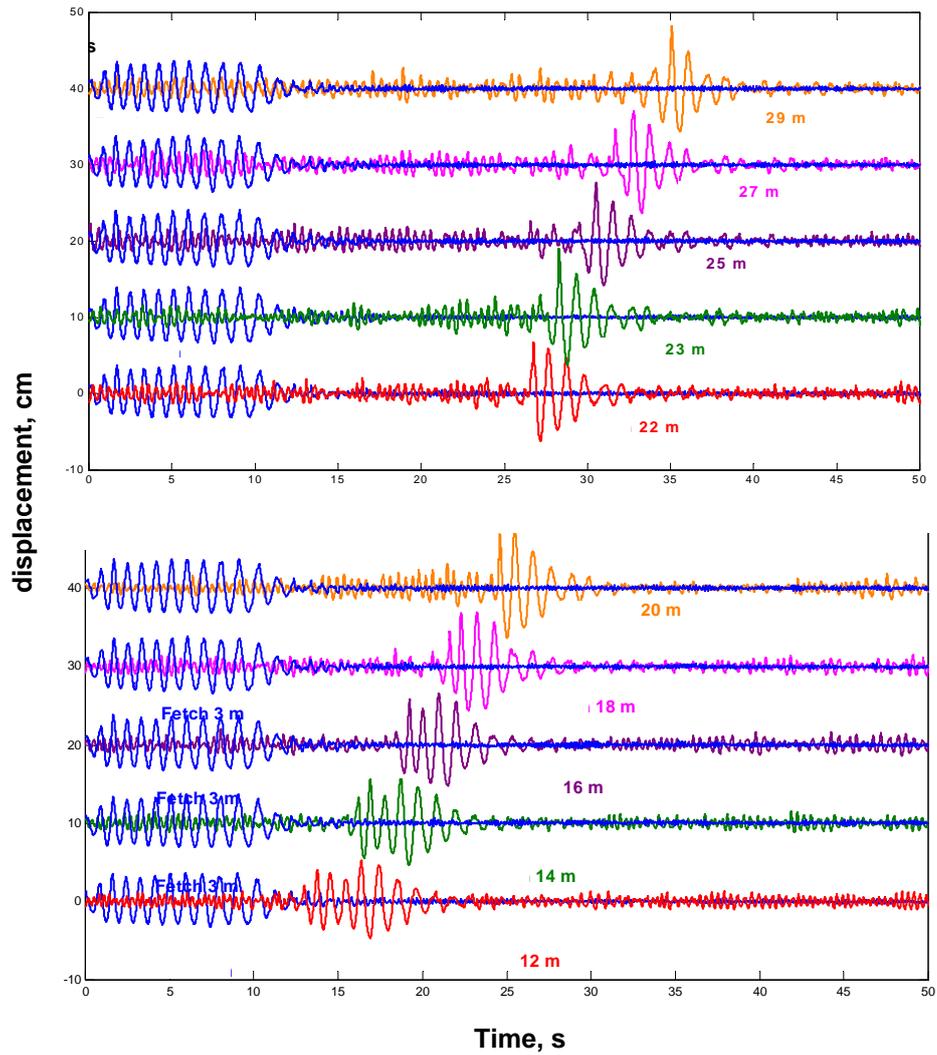

**Figure 3:** Time series of the water surface displacement, η on various distances from the wave paddle, wind speed 6 m/s; initial packet is shown on the left

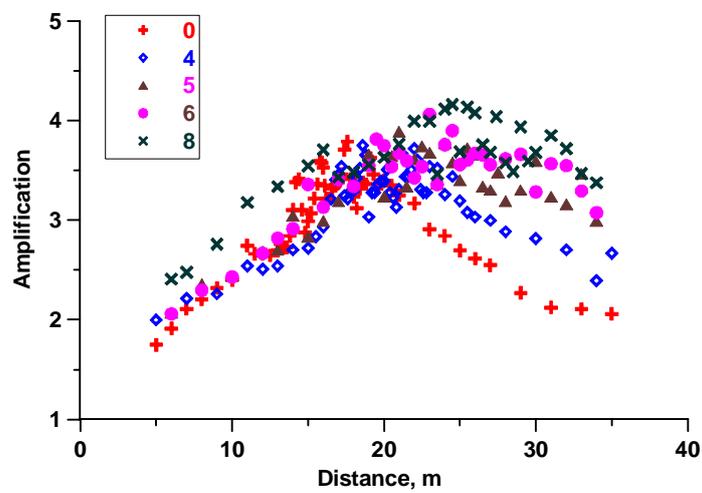

**Figure 4:** Amplification ratio versus the distance from the wave paddle for various wind speed

Results of the wavelet analysis of time series of the surface elevation at 3 m and 25 m for wind speed 8 m/s are presented in Figure 5. These figures show the time frequency evolution of the group as it propagates downstream along the wave tank. As it can be seen, the wind generated ripples have other frequencies than the transient group.

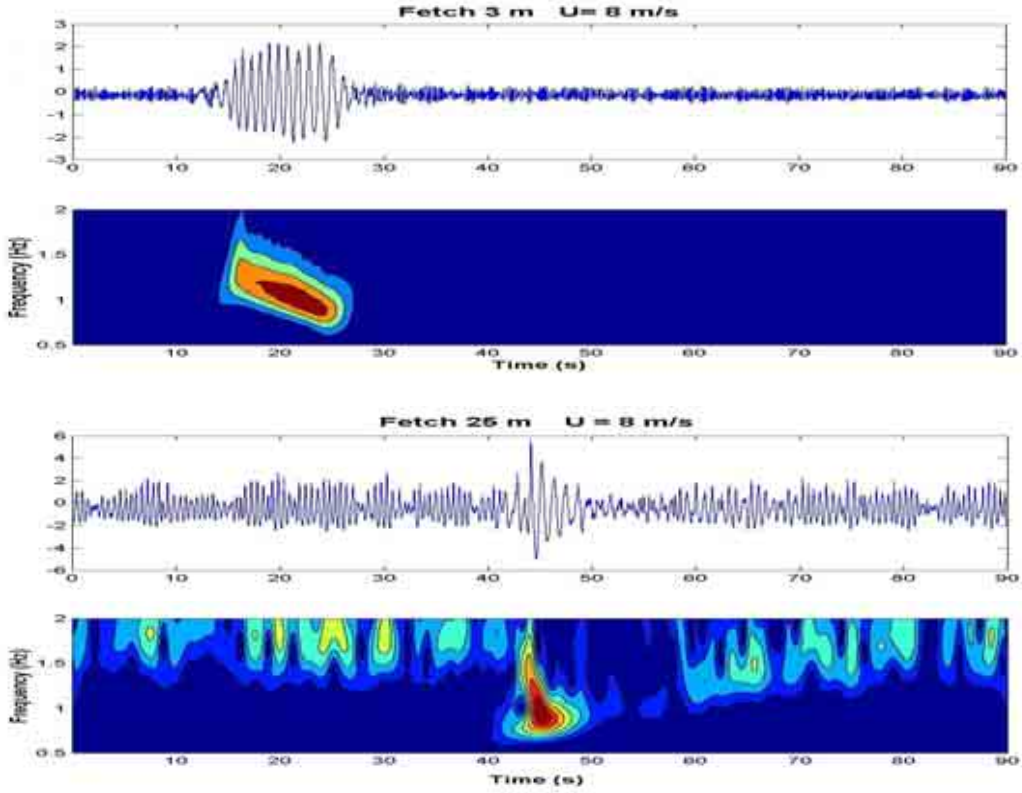

**Figure 5:** Wavelet analysis of the time records for wind speed 8 m/s

Clearly Figure 5 shows the wind waves are damped by the freak wave. This result is of the first importance for remote sensing.

**4. Focusing of narrow-banded wave groups**

The dynamics of narrow-banded linear wave packets can be described by the parabolic equation for wave amplitude

$$i\left(\frac{\partial A}{\partial t} + c_{gr}\frac{\partial A}{\partial x}\right) = \frac{\omega_0}{8k_0^2}\frac{\partial^2 A}{\partial x^2}, \quad (1)$$

where $k_0$ and $\omega_0$ are the wave number and frequency of the carrier wave, $c_{gr} = d\omega/dk$ is the group velocity. The wave amplitude, $A$, is a slowly varying function of $x$ and $t$.

Let us consider first the wave transformation of the wave packets with no wind. Transforming the variables into dimensionless form

$$\tau = \omega_0(t - x/c_{gr}), \quad y = k_0 x, \quad a = k_0 A, \qquad (2)$$

equation (1) can be re-written as spatial parabolic equation

$$i\frac{\partial a}{\partial y} = \frac{\partial^2 a}{\partial \tau^2}, \qquad (3)$$

which should be completed by the boundary condition on the paddle

$$a(\tau, y = 0) = f(\tau). \qquad (4)$$

Equation (1) or (3) are widely used to demonstrate the effect of the wave focusing of the linear transient groups (Clauss and Bergmann, 1986; Pelinovsky & Kharif, 2000; Brown & Jansen, 2001). In particular, if the boundary condition (4) corresponds to the packet with amplitude modulation of gaussian shape

$$a(\tau, 0) = A_0 \exp(-\Omega_0^2 \tau^2), \qquad (5)$$

in the process of the wave evolution its complex envelope is described by

$$a(\tau, y) = \frac{A_0}{\sqrt{1 - 4i\Omega_0^2 y}} \exp\left(-\frac{\Omega_0^2 \tau^2}{1 - 4i\Omega_0^2 y}\right), \qquad (6)$$

and, as a result, the wave has a variable amplitude and phase (frequency) modulation. The variation of the amplitude of the wave packet is

$$|a(\tau, y)| = \frac{A_0}{\left[1 + 16\Omega_0^4 y^2\right]^{1/4}} \exp\left(-\frac{\Omega_0^2 \tau^2}{1 + 16\Omega_0^4 y^2}\right). \qquad (7)$$

At each point the amplitude modulation presents a gaussian profile in time, and its peak value decreases on large distance as $y^{-1/2}$

$$|a|_{max} = \frac{A_0}{\left[1 + 16\Omega_0^4 y^2\right]^{1/4}}. \qquad (8)$$

The characteristic width of the amplitude modulation is

$$T(y) = \frac{\sqrt{1 + 16\Omega_0^4 y^2}}{\Omega_0} \qquad (9)$$

and it increases on large distances as *y*. The imaginary part of the complex amplitude gives the phase correction variable in time and space

$$\arg[a(\tau, y)] = \frac{\operatorname{atan}(4\Omega_0^2 y)}{2} - \frac{4\Omega_0^4 \tau^2 y}{1+16\Omega_0^4 y^2}. \tag{10}$$

At each point the frequency correction is

$$\Omega(\tau, y) = \frac{\partial \arg(a)}{\partial \tau} = -\frac{8\Omega_0^4 \tau y}{1+16\Omega_0^4 y^2}. \tag{11}$$

It corresponds to the linear variation of the frequency with time in the wave packet. Adding the carrier frequency, it means that the wave packet has low frequency oscillations on the front because they have large values of speed propagation (usual action of the frequency dispersion).

If this wave packet is inverted in space ($y \to -y$), it will represent a wave packet with high frequency waves on the front, which propagate slowly. Let the paddle generate such a wave packet, it means that the boundary condition (4) for complex amplitude, $a(\tau,0)$ is

$$a(\tau,0) = A_{in} \exp\left[-\Omega_{in}^2 \tau^2 + iq\Omega_{in}^2 \tau^2\right] \tag{12}$$

with three independent parameters: peak value, $A_{in}$, characteristic spectral width, $\Omega_{in}$, and phase index, *q*; formally these parameters can be obtained from (7) and (11) using $y = -L$. It means that the amplitude modulation in the process of the wave evolution from the paddle will describe by

$$|a(\tau, y)| = A_{in} \left[\frac{1+q^2}{1+q^2(y/L-1)^2}\right]^{1/4} \exp\left(-\Omega_{in}^2 \tau^2 \frac{1+q^2}{1+q^2(y/L-1)^2}\right), \tag{13}$$

where the focus distance is

$$L = \frac{q}{4\Omega_{in}^2(1+q^2)}. \tag{14}$$

At the focal point, $y = L$ the complex envelope has the real part only of the gaussian profile and it will correspond to the amplitude modulated pulse

$$a(\tau, L) = A_f \exp(-\Omega_f^2 \tau^2), \quad A_f = A_{in}\left[1+q^2\right]^{1/4}, \quad \Omega_f = \Omega_{in}\left[1+q^2\right]^{1/2}. \tag{15}$$

The solution given above describes the evolution of a free transient group with the formation of a freak wave. Results of the comparison with experimental data are presented in Figure 6. We used the following parameters: $A_f$ = 3.8 m, $f_0$ = 1 Hz and $\Omega_0$ = 0.062. In the case of no wind the agreement between theoretical and numerical results is quite good.

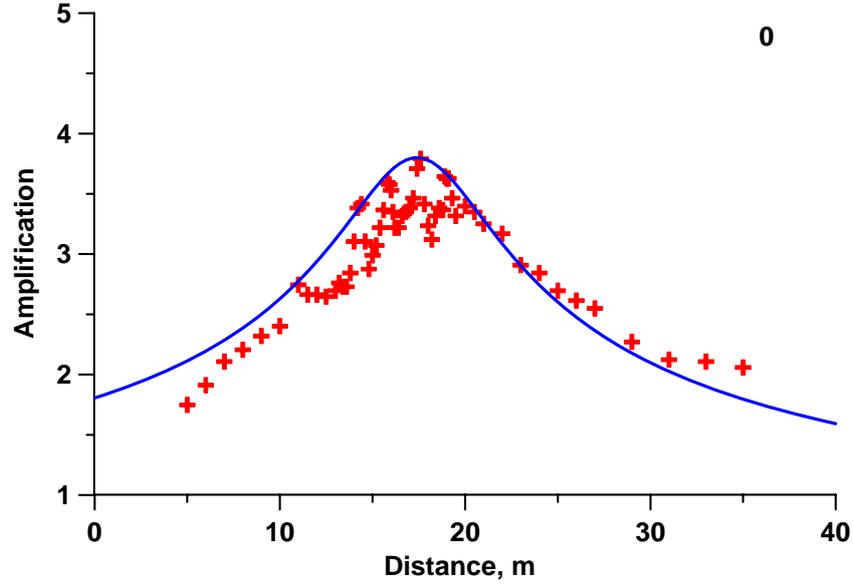

**Figure 6:** Comparison between experimental points and solution of the parabolic equation

The next step is to include the wind effect. The wind leads to the wind increment and also the induced flow. As it is known, the linear part of the any evolution equation is the inverse Fourier transformation of the linear dispersion relation. The induced flow modifies the dispersion relation which can be written as follows

$$k = \frac{g}{4u^2}\left[\sqrt{1+\frac{4u\omega}{g}} - 1\right]^2. \tag{166}$$

Then, the wave number should be developed by using Taylor series in the vicinity of the carrier frequency, $\omega_0$

$$k = k_0 + \left.\frac{dk}{d\omega}\right|_{\omega_0}(\omega-\omega_0) + \frac{1}{2}\left.\frac{d^2k}{d\omega^2}\right|_{\omega_0}(\omega-\omega_0)^2 + ..., \tag{17}$$

where $k_0$ and $\omega_0$ satisfy the dispersion relation (16), and two important coefficients in (17) equal to

$$\left.\frac{dk}{d\omega}\right|_{\omega_0} = \frac{1}{u}\left[1 - \frac{1}{\sqrt{1 + \frac{4u\omega_0}{g}}}\right], \qquad \left.\frac{d^2k}{d\omega^2}\right|_{\omega_0} = \frac{2}{g}\left(1 + \frac{4u\omega_0}{g}\right)^{-3/2} \qquad (18)$$

When the flow is absent, these coefficients transform into known formulas

$$\left.\frac{dk}{d\omega}\right|_{\omega_0} = \frac{2\omega_0}{g}, \qquad \left.\frac{d^2k}{d\omega^2}\right|_{\omega_0} = \frac{2}{g}, \qquad (19)$$

which are the inverse group velocity and inverse dispersion parameter. The evolution equation can be easily obtained from (17) considering $i(k - k_0)$ and $-i(\omega - \omega_0)$ as the differential operators $\partial/\partial x$ and $\partial/\partial t$ (in fact, coordinate and time are slow variables of the wave envelope)

$$i\frac{\partial A}{\partial x} = \frac{1}{2}\left.\frac{d^2k}{d\omega^2}\right|_{\omega_0}\frac{\partial^2 A}{\partial \tau^2}, \qquad (20)$$

where $\tau = t - x/c_{gr}$. Equation (20) reduces to (3) when the current vanishes. The wind increment can be introduced by adding a linear term

$$i\frac{\partial A}{\partial x} = \frac{1}{2}\left.\frac{d^2k}{d\omega^2}\right|_{\omega_0}\frac{\partial^2 A}{\partial \tau^2} + isA. \qquad (21)$$

This lattert term responsible of the wind generation can be eliminated by

$$A(\tau, x) = B(\tau, x)\exp(sx), \qquad (22)$$

and finally we obtain again a parabolic equation

$$i\frac{\partial B}{\partial x} = \frac{1}{2}\left.\frac{d^2k}{d\omega^2}\right|_{\omega_0}\frac{\partial^2 B}{\partial \tau^2}. \qquad (23)$$

Equation (23) was investigated earlier and we may use the previous gaussian solution (by replacing the coefficients) to study the wind effect. First of all, some conclusions can be done before any procedures. As it is evident, the wind increment modifies the wave amplitude providing additional growth of the wave field, but really this effect is not important; see Figure 5. The second one is that the group velocity ($d\omega/dk$) increases with the current and, therefore, the focal point will be shifted on large distance from the paddle. The third one is that the dispersion parameter ($d^2k/d\omega^2$) decreases with the current and, therefore, the wave amplitude will change more slowly with distance. All these conclusions correspond to the experimental data and may be illustrated by the gaussian solution again. If the wave envelope in the focal point is the gaussian pulse

$$a(\tau,0) = A_0 \exp(-\Omega^2\tau^2) \qquad (24)$$

(now we use dimensional variables), the maximal amplitude will vary with distance as

$$|a|_{max} = \frac{A_0}{\left[1 + 4(d^2k/d\omega^2)^2 \Omega^4 (x-L)^2\right]^{1/4}} . \qquad (25)$$

Figure 7 shows the comparison with observed data for wind speed value 4 m/s. Here we used experimental values for envelope in the focal point: $A_0 = 3.8$ m, $\Omega = 0.98$ Hz. The comparison is quite good.

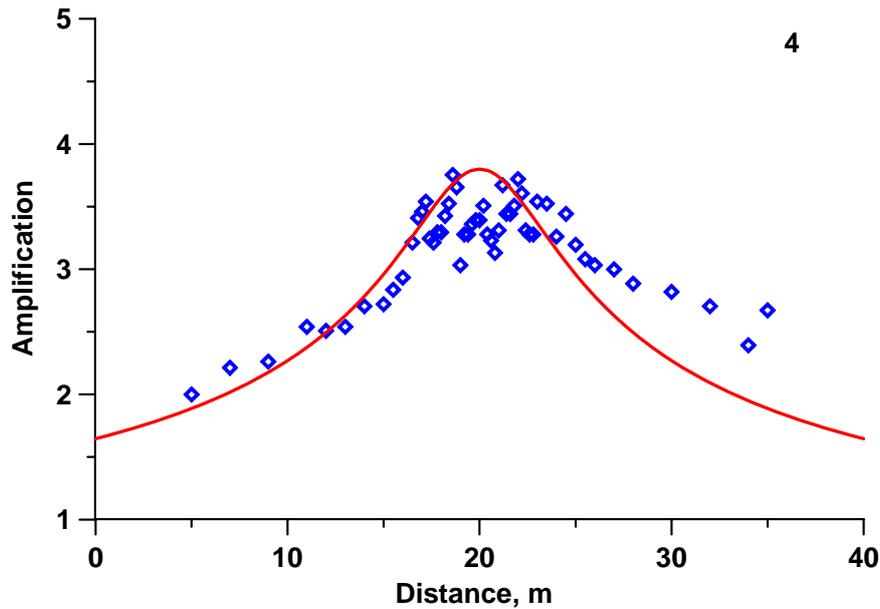

**Figure 7:** Comparison between predictions of linear parabolic equation and experimental data for wind speed 4 m/s

However, in presence of wind, the assymetry observed on the curve is not found by the model. Reul et al , 1999 have shown that over a steep wave, air flow separation process can occur inducing large local enhancement of the momentum flux from wind to the wave. We therefore could suggest that this process could occur in the same manner over the rogue wave at and downstream the focusing point. This could maintain the freak as it propagates after the focusing point.

## 5. Conclusion

The focusing of the transient wave groups under a wind flow action has been investigated experimentally and theoretically. It is shown that focal distance is increased as well as the maximum wave amplitude in this point. This effect is explained by the kinematic (Doppler) effect in the dispersion relation in the framework of the parabolic equation.

This research is supported by CNRS and for EP by FRBR (05-05-64265) and INTAS (03-514286).